\documentstyle[12pt]{article}

\begin{document}
\newcommand {\be}{\begin{equation}}
\newcommand {\ee}{\end{equation}}
\newcommand {\bea}{\begin{array}}
\newcommand {\cl}{\centerline}
\newcommand {\eea}{\end{array}}
\def\o{\over}
\def\par{\partial}
\baselineskip 0.65 cm
\begin{flushright}
IPM-97-260\\
hep-th/9712199
\end{flushright}
\begin{center}
 {\Large {\bf More on Mixed Boundary Conditions and D-branes Bound States}}
\vskip .5cm

 M.M. Sheikh-Jabbari
\footnote{ E-mail:  jabbari@theory.ipm.ac.ir}
\vskip .5cm

 {\it Institute for studies in theoretical Physics and mathematics 
IPM,

 P.O.Box 19395-5531, Tehran, Iran}\\
{\it and}
\\
{\it Department of physics Sharif University of technology, 

P.O.Box 11365-9161, Tehran, Iran}
\end{center}

\vskip 2cm
\begin{abstract}
  
In this article, applying different types of boundary conditions; Dirichlet,
Neumann, or Mixed, on open strings we realize various new brane bound 
states in string theory. 
Calculating their interactions with other D-branes, we find their charge 
densities and their tension. 
A novel feature of $(p-2,p)$ brane bound state is its "non-commutative" 
nature which is manifestly seen both in the open strings mode expansions
and in their scattering off a $D_p$-brane.
Moreover we study three or more object bound states in string theory language. 
Finally we give a M-theoretic picture of these bound states.
\end{abstract}

\newpage
\section{Introduction}
Since D-branes and their possible bound states are essential tools for 
the better understanding of various dualities, 
they have been investigated extensively [1,2,3,4,5,6,7,8,9,10].
The possible bound states of D-branes can be constructed from branes of 
the same or different dimensions.
Also, we can construct bound states of D-branes with F-strings [2,9].
Among the two object bound states, bound states of $p-$ and $p'-$ branes 
those with $p'=p+2$
are truly (non-marginally) bound states [1,3]. We use the notation 
"$(p-2,p)$ branes" for these bound states. 
For $p'=p+4$ or $p'=p$ they are marginally bounded [1,11].
Besides bound states of D-branes with themselves, bound states of 
(F-string)-($D_p$-brane) can be represented in string theory 
[7,9]. In the case of F-string-D-string bound states Witten has shown that 
they carry the charge of $U(1)$ gauge field living
in the D-string [2]. More generally in the case of other D-branes we can show 
that the branes with non-zero
electric charge (or flux) of the $U(1)$ field, form a non-marginal bound 
state of F-strings with $D_p$-brane [9].
We denote such a bound state by $(m,\ 1_p)$ brane, which carries $m$ units of 
the NSNS two form charge.

The $(m,\ 1_p)$ branes and $(p-2,p)$ branes can be represented in 
string theory by imposing {\it mixed boundary conditions} on open strings 
attached to branes [5,6,7,9,10]. 
There is a crucial difference between $(m,\ 1_p)$ and $(p-2,p)$ branes; the 
first has the "electric" flux of the $U(1)$ gauge field living in the 
D-branes [9], and the later carries the "magnetic" flux of that gauge
field [5,6,7,10].
By T-duality, a $(m,\ 1_p)$ brane is related to the moving 
$D_{(p-1)}$-brane [9]. 
In contrast the $(p-2,p)$ brane is obtained from a $D_{(p-1)}$-brane  by an 
oblique T-duality [4,9].
        
In this paper we consider the interactions of the above bound states with other
D-branes in string theory. In this way
we will find the density of various RR charges and the tension of such bound 
states and reveal the internal structure of them.

In section 2, we consider $(p-2,p)$ brane, $D_{(p-2)}$-brane interactions. 
Although the similar problem have been considered previously in the
context of SUGRA theories[4], we present string theoretic calculations. As a 
result, we find that $(p-2,p)$ brane carries
RR charge of $(p-1)$ RR form as well as the $(p+1)$ RR form charge. This 
$(p-1)$ RR form charge is homogeneously 
distributed on the surface of $D_p$-brane. The charge density of
$(p-1)$ form is exactly the same as magnetic flux of $U(1)$ gauge field, i.e., 
we have a static gas of $D_{(p-2)}$-branes on the $D_p$-brane world volume.
Since we have many $D_{(p-2)}$-branes we expect to see some non-Abelian 
gauge fields coming about [2]. This non-commutativity will be
addressed in terms of the components of open strings attached to $D_p$-brane 
with mixed boundary conditions.

In section 3, in order to investigate more on $(p-2,p)$ brane structure, we 
consider $(p-2,p)$ brane interacting with a $D_{p}$-brane.
The novel result of this interaction is that, it explicitly shows the effects 
of {\it $D_{(p-2)}$-brane} distribution in the interaction amplitude.
 
In section 4, we construct bound states of three (or more) objects in string 
theory by considering both "electric" and "magnetic" fluxes or 
"magnetic" fluxes in more than one direction. These bound states and their 
long range fields have been introduced in field theory [8]. 
Here we calculate their charge content and their tension by string theory 
methods and show that they are non-marginal bound
states of individual D-branes or F-strings with D-branes.

In section 5, we study these bound states in strong coupling regime of string 
theory i.e., we give a M-theoretic picture of such bound states.

\section {$(p-2,p)$ brane and $D_{(p-2)}$ brane Interactions}

Consider a $(p-2,p)$ brane bound state and a $D_p$-brane parallel to it. 
In this section first we realize this brane system in string theory
and then by means of their string theoretic definition we calculate the 
amplitude for the exchange of a closed string between them.

The open strings stretching between $(p-2,p)$ brane and $D_{(p-2)}$-brane 
satisfy the following boundary conditions:

\be
\sigma=0 \left\{  \begin{array}{cc}
\partial_{\sigma}X^{\mu}=0 \;\;\;\;\  \mu=0,1,...,p-2 \\
X^{\mu}=0 \;\;\;\;\ \mu=p-1,...,9 . 
\end{array}\right.
\ee
\be
\sigma=\pi \left\{  \begin{array}{cc}
\partial_{\sigma}X^{\mu}=0 \;\;\;\;\  \mu=0,1,...,p-2 \\
\partial_{\sigma}X^{p-1}+{\cal F} \partial_{\tau}X^p=0 \\
\partial_{\sigma}X^{p}-{\cal F} \partial_{\tau}X^{p-1}=0 \\
 X^{\mu}=Y^{\mu} \;\;\;\;\ \mu=p+1,...,9.  
\end{array}\right.
\ee
In the above boundary conditions $p-1,p$ components of open strings have 
{\it mixed boundary conditions} at one end and D(irichlet) boundary 
conditions at the other end. We call these components MD modes. So the above 
system is described by $p-1$ NN, 2 MD and $(9-p)$ DD modes \footnote {By 
NN and DD we mean strings their both ends satisfying Neumann and Dirichlet
boundary conditions.} and ${\cal F}$ is the "magnetic" flux of $U(1)$ gauge 
field living in $D_p$-brane.

Hence the mode expansion of quantized components are:

\be
X^{\mu}=\left\{  \begin{array}{cc}
    x^{\mu} + p^{\mu}\tau+i\sum_{n \neq 0} a_n^{\mu} {e^{-in\tau} \over n} 
\cos n\sigma \;\;\;\;\ \mu=0,..,p-2 \\
     \sum a_{n_+} {e^{-in_+\tau} \over n_+} \sin n_+\sigma +\sum a_{n_-} 
{e^{-in_-\tau} \over n_-} \sin n_-\sigma \;\;\;\;\ \mu=p-1 \\
     -i\sum  a_{n_+} {e^{-in_+\tau} \over n_+} \sin n_+\sigma +i\sum a_{n_-} 
{e^{-in_-\tau} \over n_-} \sin n_-\sigma \;\;\;\;\ \mu=p \\
      Y^{\mu} {\sigma \over \pi} +\sum_{n \neq 0} a_n^{\mu} {e^{-in\tau} 
\over n} \sin n\sigma \;\;\;\;\ \mu=p+1,..,9,
\end{array}\right.
\ee
where 
\be
n_{\pm}=n\pm {1 \over \pi} \cot^{-1}{\cal F} \;\;\;\;\ , n \in Z.
\ee 

and
\be
[a_{n_+},a_{m_-}]=n_+\delta_{n+m} \;\;\;\;\; ,\;\;\;\; [a_{n}^{\mu},a_{m}^{\nu}]
=n\delta_{n+m}\eta^{\mu\nu}.
\ee

The novel feature of the above mode expansion is that the $X^{p-1},X^p$ 
components become "non-commutative". Their non-commutativity is
controlled by ${\cal F}$, i.e., when ${\cal F}$ is zero they commute.
We will show in more detail that this non-commutativity is due to the presence 
of a distribution of parallel $D_{(p-2)}$-branes with their
world volume expanded in $0,1,...,p-2$ directions.
The corresponding $D_{(p-2)}$-brane distribution makes the $X^{p-1},X^p$ to 
become non-commuting [2].
 
Let us calculate the interactions of $(p-2,p)$ brane with a 
$D_{(p-2)}$-brane. 
This can be done by the usual techniques  [1,10,11], i.e., 
calculating the one loop vacuum graph of the open strings stretched between 
branes:

\be 
A=\int {dt \over 2t} \sum_{i,p}e^{-2\pi\alpha' t \cal H },
\ee 
where $i$ indicates the modes of the open string and $p$ their momentum,  and 
${\cal H}$ is the open string world-sheet Hamiltonian.

Along the discussions of [11] , up to tree level, the massless closed strings 
contributions are

\be
A=4V_{p-2} T_{p} T_{p-2}g^2_s {1 \over \sin \theta}(1-\cos \theta)^2 
G_{9-p}(Y^2); \;\;\;\;\; \cot\theta={\cal F}, \ee
and
\be
T_p={(4\pi^2\alpha')^{3-p} \over g_s}.
\ee
From the amplitude, the RR and NSNS potentials can be extracted [12].

\be
\left\{  \begin{array}{cc}
V_{RR}=+8V_{p-2} T_{(p-2)} T_{(p-2)}\ g^2_s\ ({{\cal F} \over \alpha'}) G_{9-p}(Y^2) \\
V_{NSNS}=-4V_{p-2} T_{(p-2)} T_{p} (1+{\cal F}^2)^{1/2}\ g^2_s\ (1+\cos^2 \theta) G_{9-p}(Y^2).
\end{array}\right.
\ee

As we see the $V_{RR}$ is proportional to ${\cal F}$. This shows that 
$({{\cal F} \over \alpha'})$ is the density of
RR $(p-1)$  form distributed on the $Dp$-brane world volume.
The NSNS potential which is due to graviton, dilaton  exchange shows that  
the mass density or tension of the  $(p-2,p)$ brane
bound state is $T_p (1+{\cal F}^2)^{1/2}$. So this bound state is 
"non-marginally" bounded.
 
Under T-duality in the $X^p$ direction the above system of branes is 
transformed to a system of two $D_{p-1}$-branes at angle 
$\theta$ [4,9].
We study the strong coupling regime of the above bound state in section 5.

\section {$(p-2,p)$ brane and $D_p$-brane Interactions}

In this section, we consider $(p-2,p)$ brane parallel to a $D_p$-brane. This 
configuration could be realized in string theory by
the following boundary conditions on open strings:
 
\be
\sigma=0 \left\{  \begin{array}{cc}
\partial_{\sigma}X^{\mu}=0 \;\;\;\;\  \mu=0,1,...,p \\
 X^{\mu}=0 \;\;\;\;\ \mu=p+1,...,9 . 
\end{array}\right.
\ee
\be
\sigma=\pi \left\{  \begin{array}{cc}
\partial_{\sigma}X^{\mu}=0 \;\;\;\;\  \mu=0,1,...,p-2 \\
\partial_{\sigma}X^{p-1}+{\cal F} \partial_{\tau}X^p=0 \\
\partial_{\sigma}X^{p}-{\cal F} \partial_{\tau}X^{p-1}=0 \\
 X^{\mu}=Y^{\mu} \;\;\;\;\ \mu=p+1,...,9.  
\end{array}\right.
\ee

These open strings have $(p-1)$ NN components, 2 MN and $(9-p)$ DD.
The mode expansions of $X^{\mu}$ are the same as (3) except for the $X^{p-1}, 
X^p$ components which are:

\be \bea{cc}
X^{p-1}= x^{p-1}+\sum a_{n_+} {e^{-in_+\tau} \over n_+} \cos n_+\sigma 
+\sum a_{n_-} {e^{-in_-\tau} \over n_-} \cos n_-\sigma  \\
X^p=x^p+\sum -i a_{n_+} {e^{-in_+\tau} \over n_+} \cos n_+\sigma 
+i\sum a_{n_-} {e^{-in_-\tau} \over n_-} \cos n_-\sigma,
\end{array}
\ee
where 
\be
n_{\pm}=n\pm {1 \over \pi} \tan^{-1}{\cal F} \;\;\;\;\ , n \in Z.
\ee 

Again $X^{p-1}, X^p$ (the MN modes) become "non-commuting". Their commutation 
relation is

\be
[X^{p-1}, X^p]=i\alpha'\sum_{n \neq 0} {1 \over n_+} \cos^2 n_+\sigma\;\;\;\ ; 
\;\;\; [x^{p-1}, x^p]={i\alpha' \over {\cal F}}.
\ee

As we see  the $X^{p-1}, X^p$ components become commuting in 
${\cal F}\rightarrow 0$ limit. The second relation tells us that
the whole $(p-1,p)$ plane is not available for the end of MN components of 
open strings and they are allowed to move in a cell
 with the area $({\alpha' \over {\cal F}})$:
\be 
\Delta x^{p-1}\Delta x^p \sim {\alpha' \over {\cal F}}.
\ee

In order to calculate the interaction of a $D_p$-brane with the above bound 
state, we use the open string one loop amplitude method.
Extracting the contributions of massless closed string we obtain:

\be \bea{cc}
A=4V_{p-1} \alpha' T_{p}^2 g^2_s \ {1 \over \sin \theta}(1-\cos \theta)^2 
G_{9-p}(Y^2)\\
\;\;\;\;\;\;\;\;\;  =4V_{p-1} ({\alpha' \over {\cal F}})({T_{p} \over 
\cos \theta}) T_p g^2_s(1-\cos \theta)^2 G_{9-p}(Y^2),
\eea
\ee
where  ${\cal F}=\tan \theta$. The contributions of RR and NSNS closed strings 
are

\be
\left\{  \begin{array}{cc}
V_{RR}=+8V_{p-1} ({\alpha' \over {\cal F}})T_p^2  \ g^2_s\ G_{9-p}(Y^2) \\
V_{NSNS}=-4V_{p-1} ({\alpha' \over {\cal F}})T_p \ T_p (1+{\cal F}^2)^{1/2}\ 
g^2_s\  (1+\cos^2 \theta) G_{9-p}(Y^2).
\end{array}\right.
\ee

As we expected, instead of the usual $D_p$-brane world volume factor 
$(V_{p+1})$, we obtain $(V_{p-1}\times {\alpha' \over {\cal F}})$.
This shows that the end of open  strings is limited to move in the cell of 
equation (15).

The RR contribution is proportional to $T_p^2$, showing that the $(p-2,p)$ 
brane carries the {\it unit} charge density of the RR $(p+1)$ form.

The NSNS contributions again justifies that the tension of the corresponding 
bound state is $T_p\times (1+{\cal F}^2)^{1/2}$.

It is worth noting that applying T-duality two times on the mixed 
directions 
of the branes system of this section, we find brane configuration
of the previous section. More precisely, the system of $(p-2,p)$ brane 
parallel to a $D_p$-brane, under such a T-duality transforms to $(p-2,p)'$ 
brane parallel to a $D_(p-2)$-brane, where if $(p-2,p)$ brane is associated 
with magnetic flux ${\cal F}$, the $(p-2,p)'$ brane is associated with 
${-1 \over {\cal F}}$. The  relation between related fluxes are directly seen 
from the corresponding interaction amplitudes.

Altogether we find that $(p-2,p)$ brane is an object with unit charge of RR 
$(p+1)$ form, the charge density equal to $({{\cal F} \over \alpha'})$
for the RR $(p-1)$ form and the mass density $T_p\times (1+{\cal F}^2)^{1/2}$.
Our interpretation for its internal structure is that parallel 
$D_{(p-2)}$-branes have been distributed homogeneously in the $D_p$-brane and 
hence two of the internal directions of the $(p-2,p)$ brane which are 
normal to $(p-2)$ branes world volume, show "non-commutative" effects.

\section { Bound States of Three or More Objects}

We can build bound states of D-branes containing  $p,p-2,p-4,...$  branes [10] 
and also the bound states of F-strings with  $p,p-2,p-4,..$ branes,
in the same spirit we have shown in previous sections. Let us consider a 
general magnetic flux of $U(1)$ gauge field living in a $D_p$-brane. 
By imposing  these non-vanishing magnetic fluxes ($F_{ij}$) on boundary 
condition of the open strings attached to brane, 
we can construct a more general branes bound state.
For every non-zero $F_{ij}$ we have a distribution of $D_{(p-2)}$-branes in 
$(i,j)$ plane. For example if we have $F_{12}$,$F_{34}$ components only, our
bound state consists of distributions of $D_{(p-2)}$-branes in $(1,2)$ and 
$(3,4)$ planes and also a distribution of $D_{(p-4)}$-branes  in the 
world volume of the $D_{p-2}$-branes in $(3,4)$ and $(1,2)$ planes respectively. 
So we have a bound state of three objects: $(p,p-2,p-4)$ brane, which
corresponding tension and charge densities are

\be
\bea {cc}
T=T_p (1+{\cal F}_{12}^2)^{1/2}(1+{\cal F}_{34}^2)^{1/2} \\
{\rm RR\ (p+1)\ form\ charge\ density} = T_p \\
{\rm RR\ (p-1)\ form\ charge\ density} = T_{p-2}({{\cal F}_{12} \over \alpha'})
\;\;\;\ \ and \;\;\;\ T_{p-2}({{\cal F}_{34} \over \alpha'})  \\
{\rm RR\ (p-3)\ form\ charge\ density} =T_{p-4}({{\cal F}_{12} \over \alpha'}) 
({{\cal F}_{34} \over \alpha'}).
\eea
\ee

These charge densities could be directly checked in brane interactions.

One can consider cases which also include the "electric" flux. In these cases 
we find a bound state of F-string with various D-branes. In the corresponding
bound state the NSNS two form charge density is proportional to 
$(electric\ flux \times \alpha'^{3-p \over 2})$. It is worth to note that the 
NSNS two form charge,
unlike the magnetic flux case, is independent of string coupling constant [9].

\section{Strong Coupling Limit}

In this part, we study the strong coupling regime of the $(m,\ 1_p)$ and 
$(p-2,p)$ branes.

{\it even p case}

$p=0$

In this case because  the world volume of the $D_0$-brane is just one 
dimensional, we have not such bound states.

$p=2$
 
$i$) $(0,2)$ brane: As we know $D_2$-branes at strong coupling are $M_2$-branes
of M-theory and $D_0$-branes are gravitational waves of 11 dimensional SUGRA.
Hence we expect that $(0,2)$ brane is just kinematically related to $M_2$-branes
[13] \footnote{I am grateful to B. Pioline for his comments on 
this subject.}. Let us consider a {\it moving} $M_2$-brane. If we compactify  
the velocity direction then the three form of the eleven dimensional
SUGRA gives a "magnetic" flux of the two form living in the $D_2$-brane.
The strength of this flux is proportional to the brane velocity.

The tension of a moving membrane after doing the above dimensional reduction is
\be
T^2=m_p^6(1+ (Rv)^2)\;\;\;\; ; \;\;\; v=n/R.
\ee

So in type IIA language:
\be
T^2=T_2\ ^2(1+ {\cal F}^2)^2,
\ee
where ${\cal F}$ is the magnetics flux and $T_2={m_s^3 \over g_s}=m_p^3$.

$ii$) $(m,\ 1_2)$ brane: This bound state in M-theory is just a $M_2$-brane 
compactified on a direction making an angle with the membrane.
The magnitude of NSNS charge is proportional to the $\cos \theta$, $\theta$ 
is the corresponding angle.

$p=4$

$i$) $(2,4)$ brane: The type IIA $D_4$-branes at strong couplings are wrapped 
$M_5$-branes. So the $(2,4)$ brane at strong couplings
is a "non-marginal" bound state of $M_2$- and $M_5$-branes [14], where the 
world volume of the $M_2$-brane in bound state is not 
wrapped around the compact direction.

$ii$) $(m,\ 1_4)$ brane: This brane is just the $M_2$, $M_5$-brane bound state
where the $M_2$-brane is $m$ times wrapped around
the compact direction.

$iii$) $(m,2,4)$ brane: This bound state is the $M_2$, $M_5$-brane bound state 
in which the corresponding membrane makes an angle with the compact direction.
         
{\it odd p case}

In this case the Sl(2,Z) symmetry of IIB theory determines the string coupling 
picture of the branes bound states.

$p=1$

This case corresponds to $(m,\ n)$ strings of IIB theory. $(m,\ n)$ strings 
have been studied extensively both in field theory language [2,15]
and in string theory [9]. These strings are related to the moving $D_0$-branes 
of type IIA by T-duality [9].
 
$p=3$

In this case $(m,\ 1_3)$ and $(1,3)$ brane bound states are related by S-duality
of the (3+1) dimensional YM theory
living in $D_3$-brane world volume. This S-duality is a part of Sl(2,Z) 
symmetry of IIB theory projected on $D_3$-brane.

The $(F-,D- string,D_3)$ brane bound states also can appear. They  are also 
related to  $(m,\ 1_3)$ branes by Sl(2,Z) duality.
The $(m,\ 1_3)$ brane itself is T-dual to a moving $D_2$-brane of IIA theory.

$p=5$

As we know $D_5$-branes under a "S" transformation of Sl(2,Z) group transform 
to $NS_5$-branes. Although $NS_5$-branes are not properly understood in usual 
string theory limit, we expect at strong couplings the $(m,\ 1_5)$ brane  
become a ($NS_5$-brane)-(D-string) bound state or more generally the 
($(m,n)$ string)-($D_5$-brane) bound state
become a bound state of ($(n,-m)$ string)-($NS_5$-brane). These bound states 
and also the bound state of five-branes with three-branes are not fully 
understood in terms of $(5+1)$ dimensional field theory living in 
five-branes.


{\bf Acknowledgements}

The author would like to thank H. Arfaei who had contribution at the early 
stages of this work. I am  grateful to A.H. Fathollahi for his useful 
discussions. I am grateful the A. Tseytlin and P. Townsend for reading the 
manuscript.
I would also like to thank theory division of CERN where this work completed.

\end{document}